\documentclass[aps,pre,twocolumn,showpacs,superscriptaddress]{revtex4}
\usepackage{amsmath}
\usepackage{graphicx,color,epsfig}
 
\begin{document}

\title{Stratified economic exchange on networks}
\author{J. L. Herrera}
 \affiliation{Escuela B\'asica de Ingenier\'ia, Universidad de Los Andes, M\'erida, Venezuela}
\author{M. G. Cosenza}
\affiliation{Centro de F\'isica Fundamental, Universidad de Los Andes, M\'erida, M\'erida 5251, Venezuela}
\author{K. Tucci}
\affiliation{SUMA-CeSiMo, Universidad de Los Andes, M\'erida, Venezuela}
\date{\today}

\begin{abstract}
We investigate a model of stratified economic interactions between agents when the notion of spatial location is introduced. The agents are placed on a network with near-neighbor connections. Interactions between neighbors can occur only if the difference in their wealth is less than a threshold value that defines the width of the economic classes.
By employing concepts from spatiotemporal dynamical systems, three types of patterns can be identified in the system as 
parameters are varied: laminar, intermittent and turbulent states.  
The transition from the laminar state to the turbulent state is characterized by the activity of the system, a quantity that measures the average exchange of wealth over long times.  The degree of inequality in the wealth distribution for different parameter values is characterized by the Gini coefficient. High levels of activity are associated to low values of the Gini coefficient. It is found that
the topological properties of the network have little effect on the activity of the system, but the Gini coefficient increases when the clustering coefficient of the network is increased.
\end{abstract}
\maketitle

\section{Introduction}
Social stratification refers to the classification of individuals into groups or classes based on shared socio-economic or power conditions within a society \cite{Grusky}. A characteristic feature of stratified societies is that individuals tend to interact more strongly with others in their own group. This tendency has been observed in class endogamy \cite{Belding}, scientific communities and citations \cite{Lehman}, population biology \cite{Vasquez}, human capital \cite{Martins}, opinion formation \cite{Martin}, epidemic dynamics \cite{Masuda}, and economic exchanges between banks \cite{Inaoka}. Recently, the effects of social stratification on the wealth distribution of a system of interacting economic agents have been studied \cite{Laguna}. In this model, agents behave as particles in a gas and they can interact with each other at random, as in most models that have been proposed for economic exchange \cite{Yakovenko,Chattarjee,Slanina}. However, many real social and economic systems can be described as complex networks, such as small-world networks and scale-free networks \cite{Watts,Barabasi,Newman}. Some models have considered economic dynamics on networks; for example, Refs.~\cite{Pianegonda1} and \cite{Garlaschelli} studied the effects of the network topology on wealth distributions; while Ref.~\cite{Ausloos} proposed a model of closed market on a fixed network with free flow of goods and money.

In this paper, we study the effects of the topology of a network on the collective behavior of a system subject to stratified economic exchanges. Our model, based on the interaction dynamics in a stratified society proposed by Laguna et al. \cite{Laguna}, is presented in Sec.~2. The inclusion of a spatial support allows to employ concepts from the dynamics of spatiotemporal systems in economic systems. Our results indicate that the size of the local neighborhood plays an important role for achieving an equitable distribution of wealth in systems possessing stratified economic exchange. Conclusions are presented in Sec.~3.

\section{The Model}
We consider a network defined by following the algorithm of construction of small-world networks originally proposed by Watts and Strogatz \cite{Watts}. We start from a regular ring with $N$ nodes, where each node is connected to its $k$ nearest neighbors, $k$ being an even number. Then, each connection is
rewired at random with probability $p$ to any other node in the network. After the rewiring process, the number of elements coupled
to each node -- which we call neighbors of that node -- may vary,
but the total number of links in the network is constant and
equal to $Nk/2$. The condition $\log N \leq k \leq N$ is employed
to ensure that no node is isolated after the rewiring process, which results in
a connected graph. For $p = 0$, the network corresponds to a regular ring, while for $p = 1$ the resulting
network is completely random. With this algorithm, a small-world network is formed for values of the probability in the intermediate range  \cite{Watts}. A small-world network is characterized by a high degree of clustering, as in a regular lattice, and a small characteristic path length compared to the size of the system. 

We consider a population of $N$ interacting agents placed at the nodes of this network.
At a discrete time $t$, an agent $i$ ($i=1,\ldots,N$), is characterized by a wealth $w_i(t)\geq 0$ and a fixed risk aversion factor $\beta_i$, where the values $\beta_i$ are randomly and uniformly distributed in the interval $[0,1]$. The quantity $(1-\beta_i)$ measures the fraction of wealth that agent $i$ is willing to risk in an economic interaction \cite{Chattarjee,Chakrabarti,Iglesias}. The initial values $w_i(0)$ are uniformly distributed at random in the interval $w_i(0) \in [0,W]$. We assume that the total wealth of the system, $W_T=\sum_iw_t(i)$, is conserved.

For simplicity, we assume that the stratification of economic classes is uniform, i.e., all classes have the same width, denoted by a parameter $u$. Thus, agents $i$ and $j$ belong to the same economic class if they satisfy the condition 
$|w_i(t) - w_j(t)| < u$. Stratified economic exchange means that only agents belonging to the same economic class may interact. As a consequence of these interactions, the wealth of the agents in the system will change. At each time step $t$, the dynamics of the system is defined by iterating the following steps:

\begin{enumerate}
\item Choose an agent $i$ at random.
\item Choose randomly an agent $j \neq i$ from the set of neighbors of agent $i$, i.e., $j \in [ i-k/2, i+k/2 ]$.
\item Check if they belong to the same economic class, i.e.,
\begin{equation}
|w_i(t) - w_j(t)| < u.
\end{equation}
Repeat steps (1) and (2) until condition (3) is achieved. 
\item Compute the amount of wealth $\Delta w(t)$ to be exchanged between agents $i$ and $j$, defined as
\begin{equation}
\Delta w(t) = \mbox{min}[(1-\beta_i)w_i(t);(1-\beta_j)w_j(t)].
\end{equation}
\item Calculate the probability $r$ of favoring the agent that has less wealth between $i$ and $j$ at time $t$, defined as
\cite{Laguna,Iglesias}
\begin{equation}
r=\frac{1}{2} + f \times \frac{|w_i(t) - w_j(t)|}{w_i(t) + w_j(t)},
\end{equation}
where the parameter $f \in [0,1/2]$.
\item
Assign the quantity $\Delta w(t)$ with probability $r$ to the agent having less wealth and with probability $(1-r)$ to the agent with greater wealth between $i$ and $j$.
\end{enumerate}

The parameter $f$ describes the probability of favoring the poorer of the two agents when they interact.
For $f=0$ both agents have equal probability of receiving the amount $\Delta w(t)$ in the exchange, while for $f=1/2$ the agent with less wealth has the highest probability of receiving this amount. In a typical simulation following 
these dynamical rules, and after a transient time, this dynamical network reaches a stationary state where the total wealth $W_T$ has been redistributed between the agents.

\begin{figure}[h]
\begin{center}
\includegraphics[scale=0.17,angle=0]{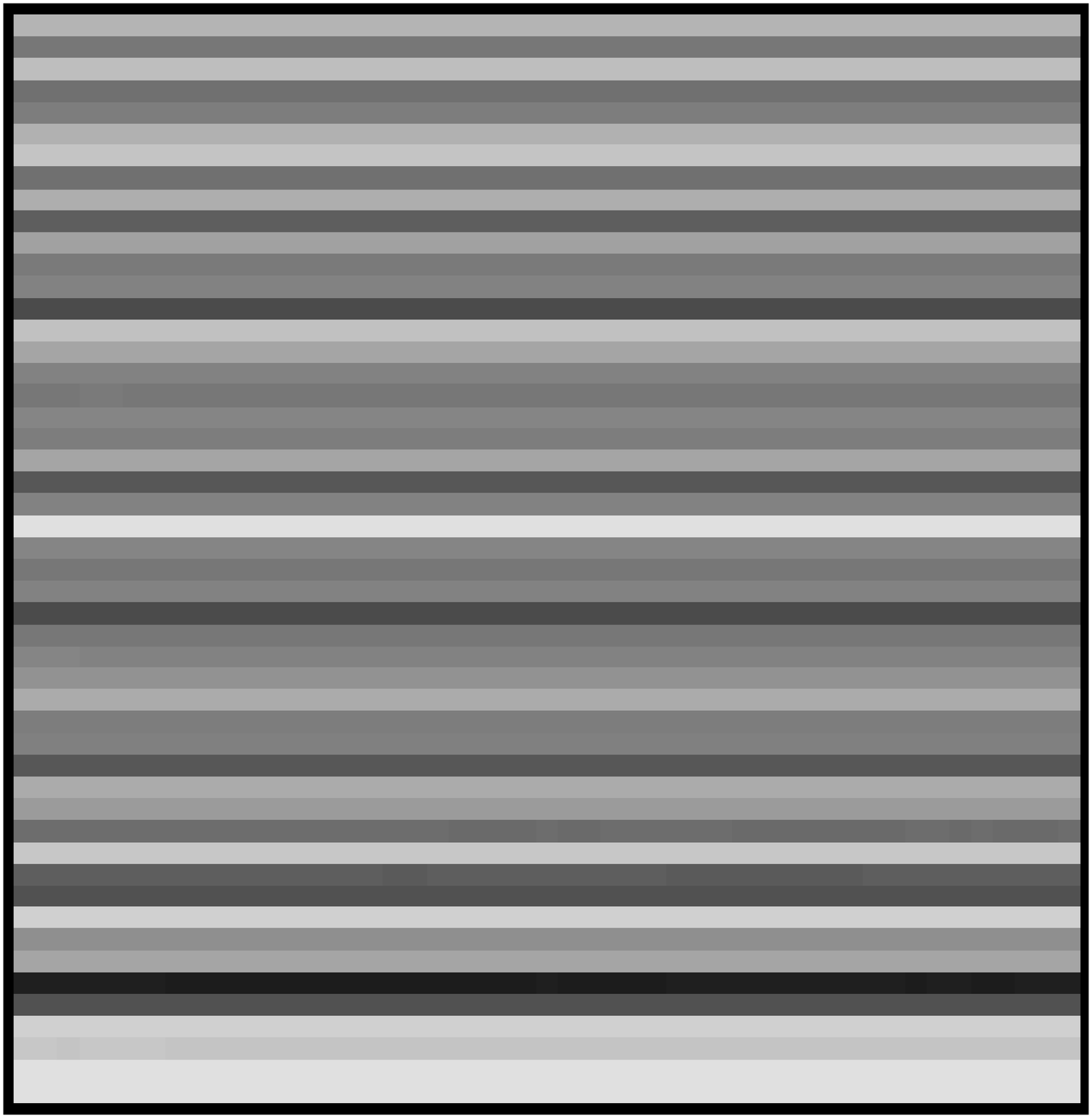}

\vspace{0.1cm}

\includegraphics[scale=0.17,angle=0]{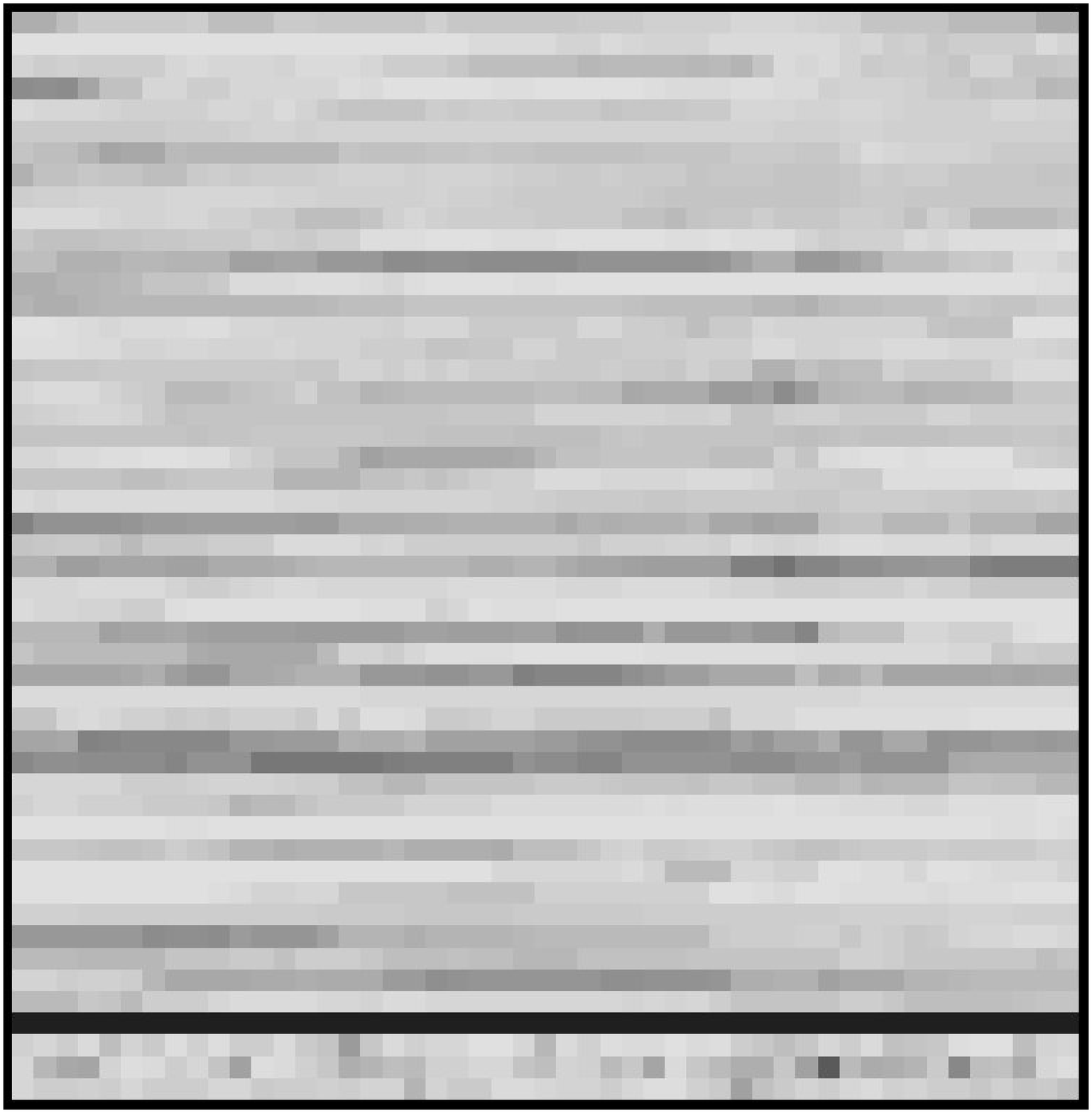}

\vspace{0.1cm}

\includegraphics[scale=0.17,angle=0]{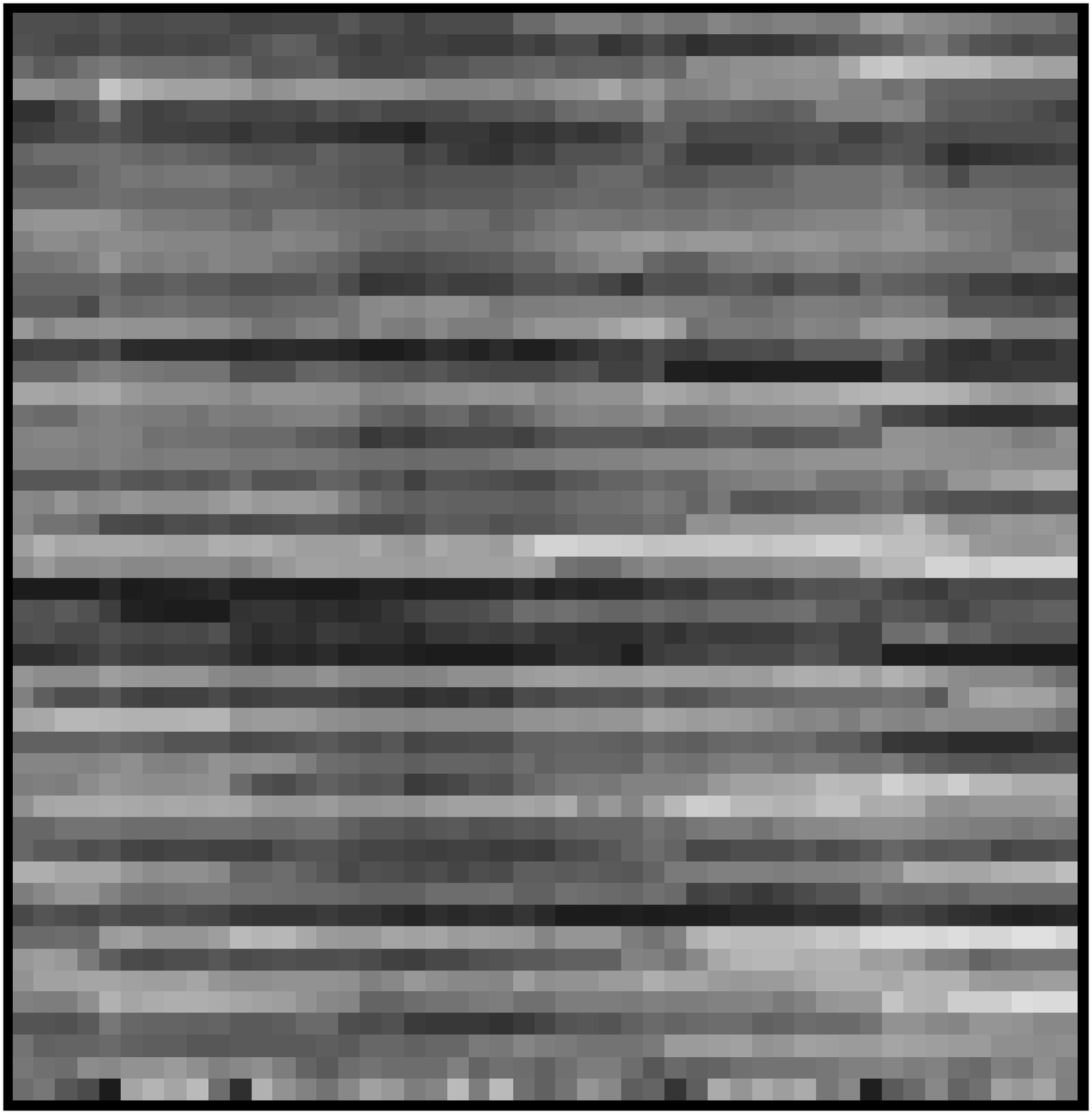}
\end{center}
\caption{Spatiotemporal patterns in a one-dimensional lattice with $k=2$, size $N=50$ and $W=1$, after discarding
$5000$ time steps. The vertical axis describes the ordered position $i$ of the agents in the lattice, increasing  from bottom to top.  Horizontal axis represents time, increasing from left to right. The wealths $w_i(t)$ evolving in time are represented by a color code. The color palette goes from light gray (the poorest agent) to dark gray (the richest agent). Top: laminar state; $u=10$, $f=0.001$. Center: spatiotemporal intermittent state; $u=3, f=0.4$. Bottom: turbulent state; $u=30$, $f=0.4$.}
\label{fig1}
\end{figure}

The spatial localization of the interacting economic agents allows to see this system as a spatiotemporal dynamical system.
Figure~\ref{fig1} shows the spatiotemporal patterns of wealth arising in a network with $k=2$ and $p=0$, corresponding 
to a regular one-dimensional lattice with periodic boundary conditions, for different values of the parameters. In analogy to many nonlinear spatiotemporal dynamical systems \cite{Manneville}, this network of economic agents can exhibit three basic states depending on parameter values: a stationary, coherent or laminar state (left panel), where the wealth of each agent $i$ maintains a constant value; an intermittent state (center panel), 
characterized by the coexistence of coherent and irregular domains evolving in space and time; and a turbulent state (right panel) where the wealth values change irregularly in both space and time.

To characterize the transition from the laminar to the turbulent state, via spatiotemporal intermittency, we employ the average wealth exchange for long times, a quantity that we call the activity of the system and define as
\begin{equation}
\label{activity}
A=\frac{1}{T-\tau}\sum_{t=\tau}^{T} \Delta w(t),
\end{equation}
where $\tau$ is a transient number of steps that are discarded before taking the average. The laminar phase is associated to values $A=0$, where no transactions take place in the asymptotic state of system, while the turbulent phase is characterized by $A >0$. 

\begin{figure}[h]
\begin{center}
\includegraphics[scale=0.39,angle=90]{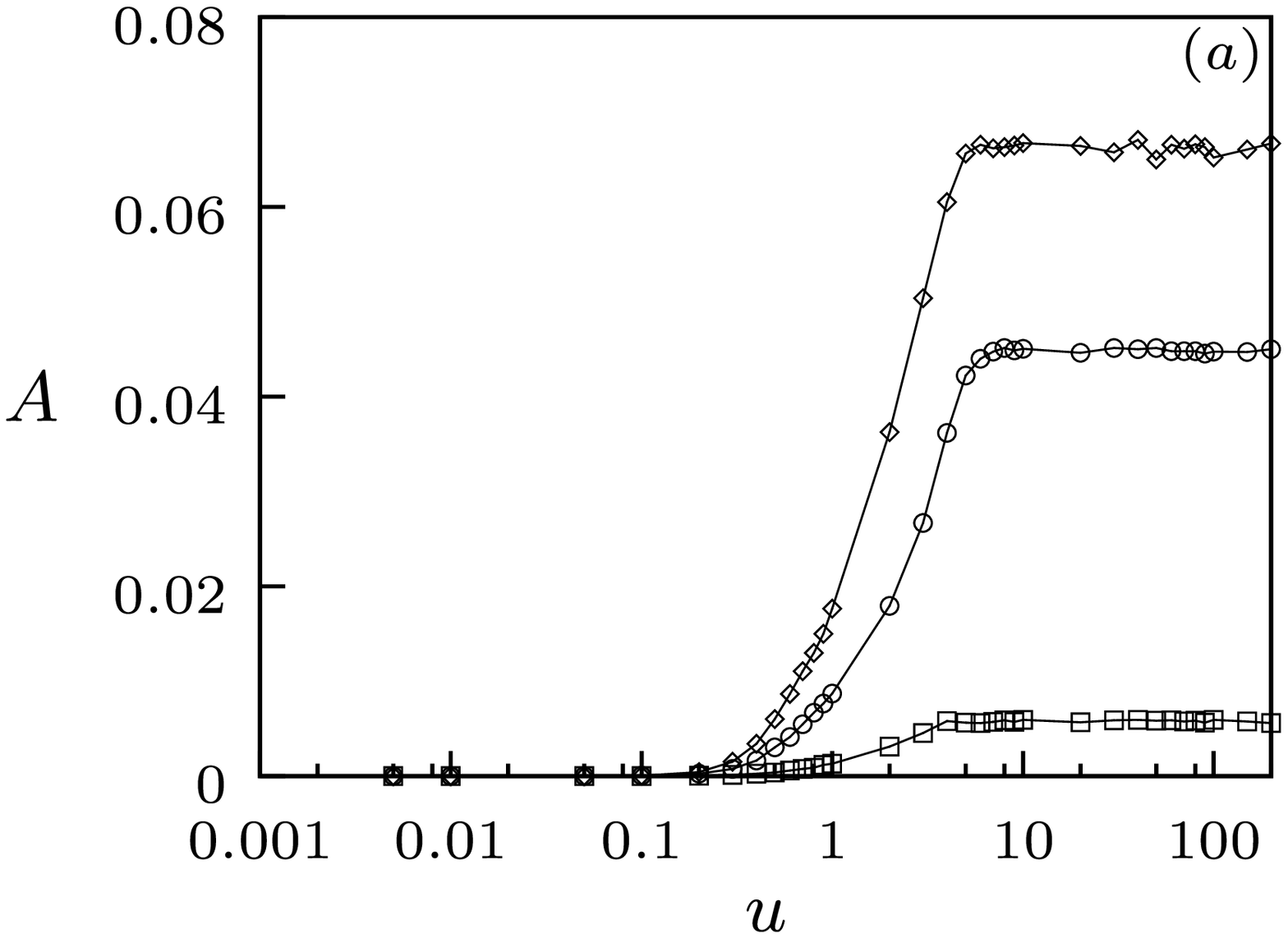}

\vspace{0.3cm}

\includegraphics[scale=0.39,angle=90]{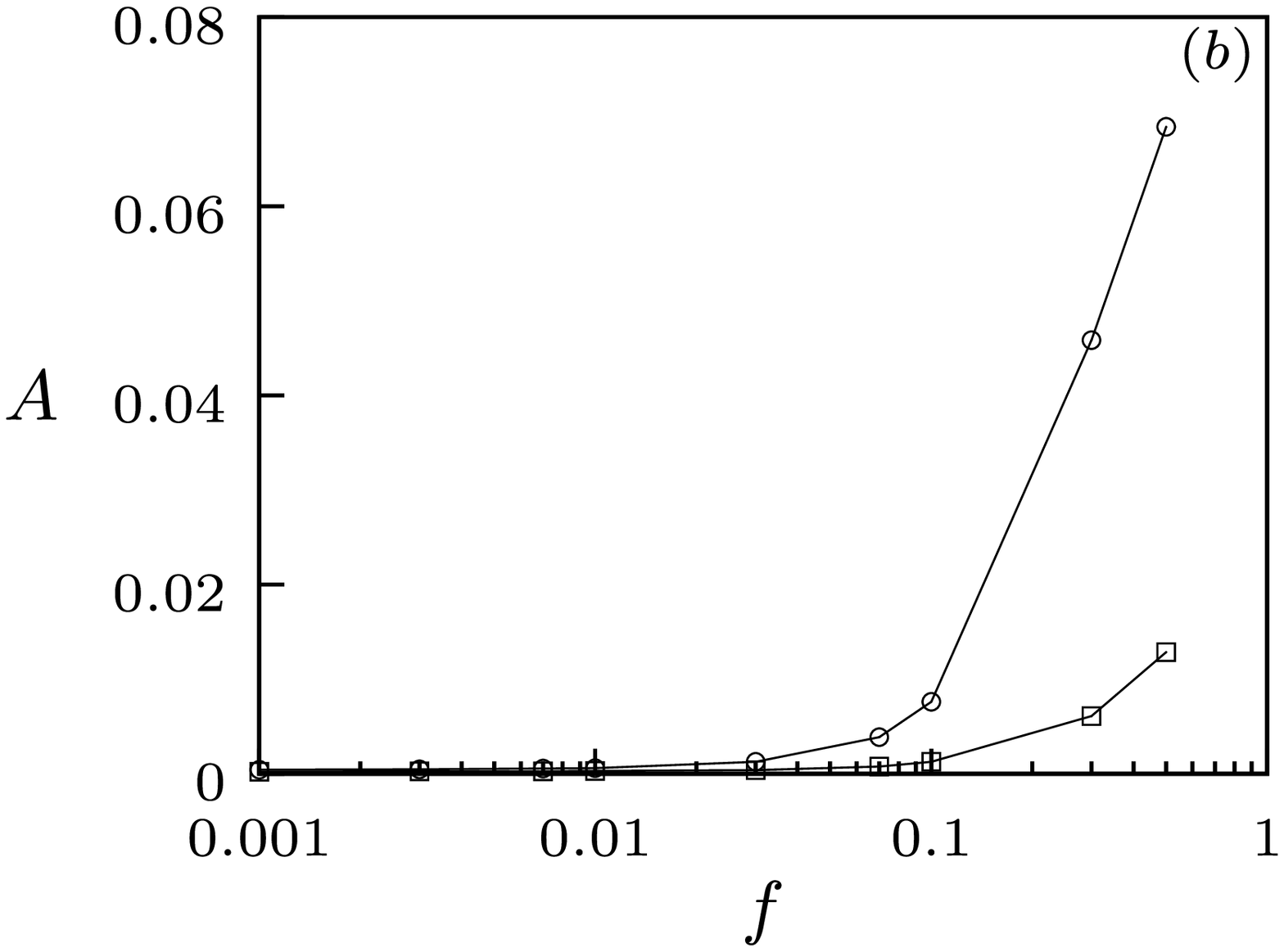}
\end{center}
\caption{(a) Activity as a function of $u$ in a regular lattice ($p=0)$ with fixed $k=4$, for different values of $f$. The curves correspond to $f=0.5$ (diamonds); $f=0.3$ (circles); and $f=0.1$ (squares).(b) Activity as a function of $f$ in a regular lattice  with $k=4$, for $u=1$ (squares), and $u=30$ (circles).}
\label{fig2}
\end{figure}

In our calculations, we have fixed these values of parameters: size $N=10^4$, $\tau=10^8$, $T=2\times 10^4$, and $W=1$. Each value of the statistical quantities shown has been averaged over $100$ realizations of initial conditions.
Figure~\ref{fig2}(a) shows the activity in the system as a function of the width of the economic classes $u$ for different values of the parameter $f$. The transition from the laminar phase to the turbulent state occurs about the value $u\approx W=1$ in all cases. When the value of the width $u$ reaches the value of the maximum initial wealth of the agents, exchanges may take place in every neighborhood, and this is reflected in the increase in the activity in the system. For $u > W$, interactions continue to occur in the entire system and the total wealth exchanged reaches the maximum amount allowed by the favoring parameter $f$. 
Thus, the activity in the system reaches an almost constant value in this region, for a given value of $f$. 
On the other hand, Figure~\ref{fig2}(b) shows the activity in the system as function of $f$.
The increment in $f$ enhances the transfer of wealth from richer to poorer agents. Therefore, the
probability that neighboring agents belong to the same economic class increases, and so does the probability that they exchange wealth. As a consequence, the activity in the system increases with increasing $f$.

\begin{figure}[h]
\begin{center}
\includegraphics[scale=0.39,angle=90]{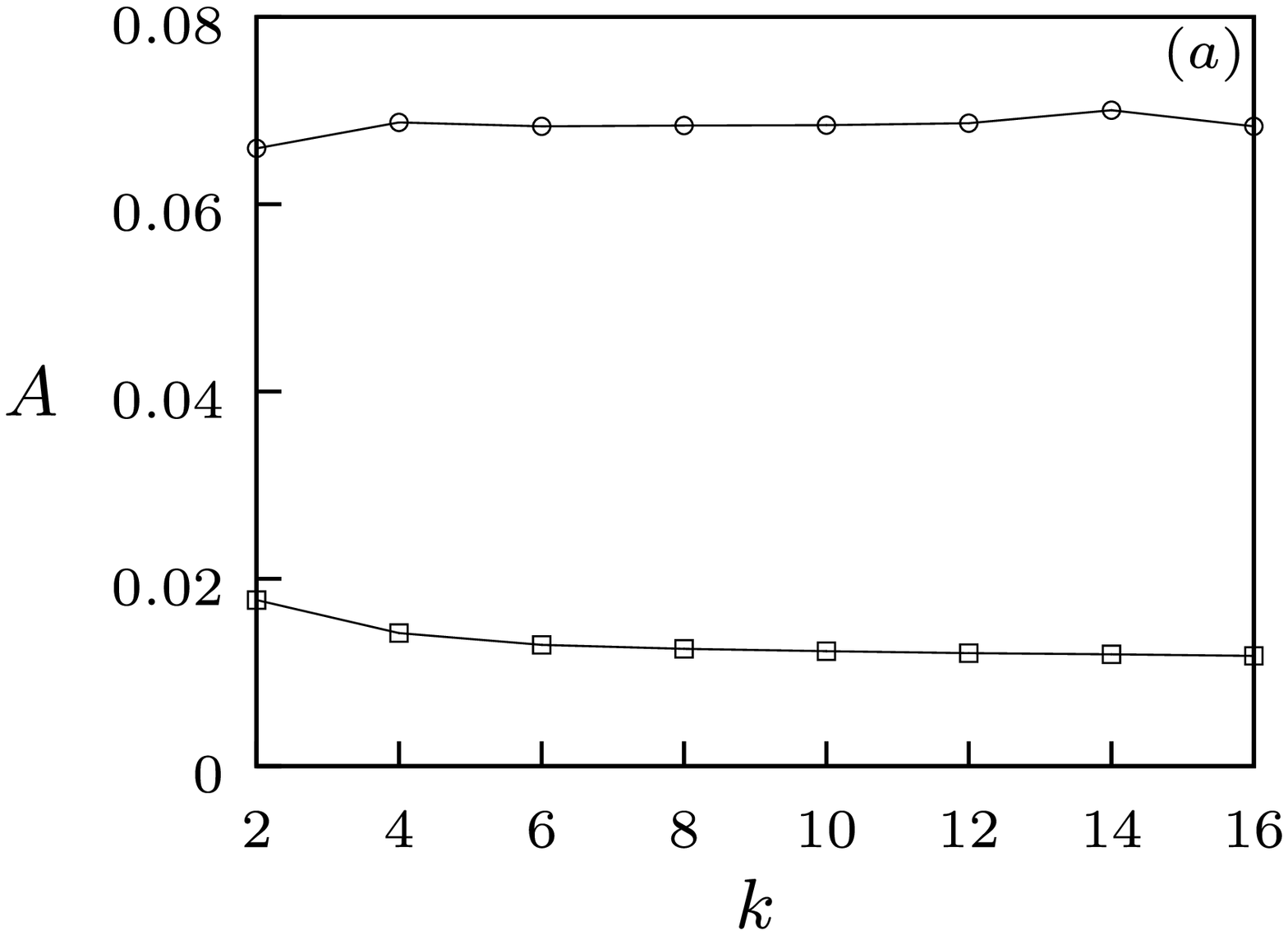}

\vspace{0.3cm}

\includegraphics[scale=0.39,angle=90]{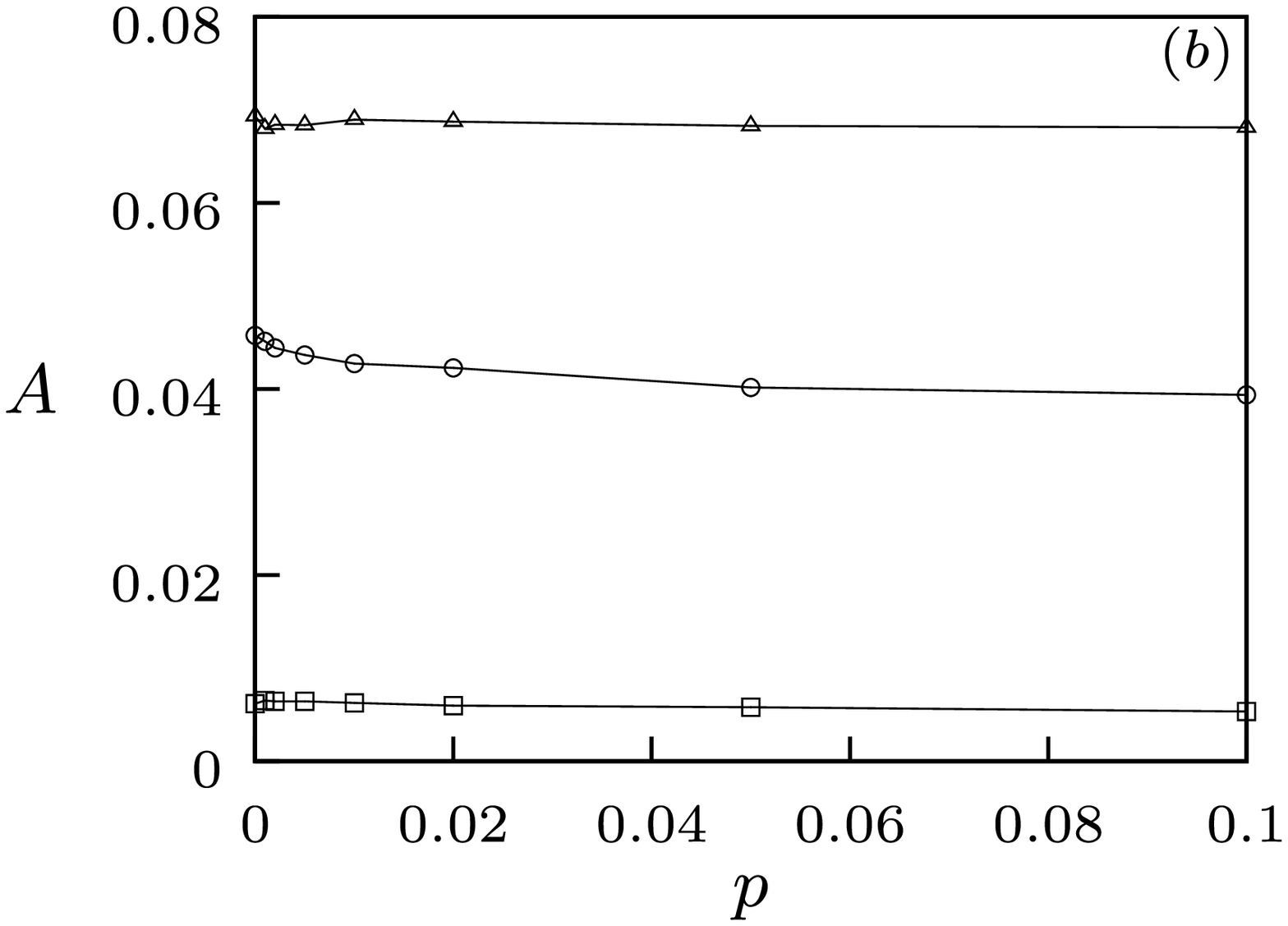}
\end{center}
\caption{(a) Activity as a function of $k$, on a regular lattice with $p=0$, fixed $f=0.5$, and for different values $u=1$ (squares) and $u=10$ (circles).
(b) Activity as a function of $p$, on a network with $k=4$, and for different values $f=0.5$ (triangles), $f=0.3$ (circles) and $f=0.1$ (squares).}
\label{fig3}
\end{figure}

To explore the effects of the network topology on the collective properties of the system, we show in Figure~\ref{fig3}(a) the activity as a function of the size of the neighborhood $k$ in the network, for different values of $u$. The range of the local interaction, given by $k$, has little effect on the activity. Similarly, Figure~\ref{fig3}(b) shows the activity as a function of
the rewiring probability in the network, for fixed $k=4$. We see that the exchange activity in the system is practically unaffected by the topological properties of the network, represented by $k$ and $p$. Thus, the parameters of the dynamics, $f$ and $u$, are more relevant for the increase in the activity in the system than the topological parameters of the underlying network.

\begin{figure}[h]
\begin{center}
\includegraphics[scale=0.39,angle=90]{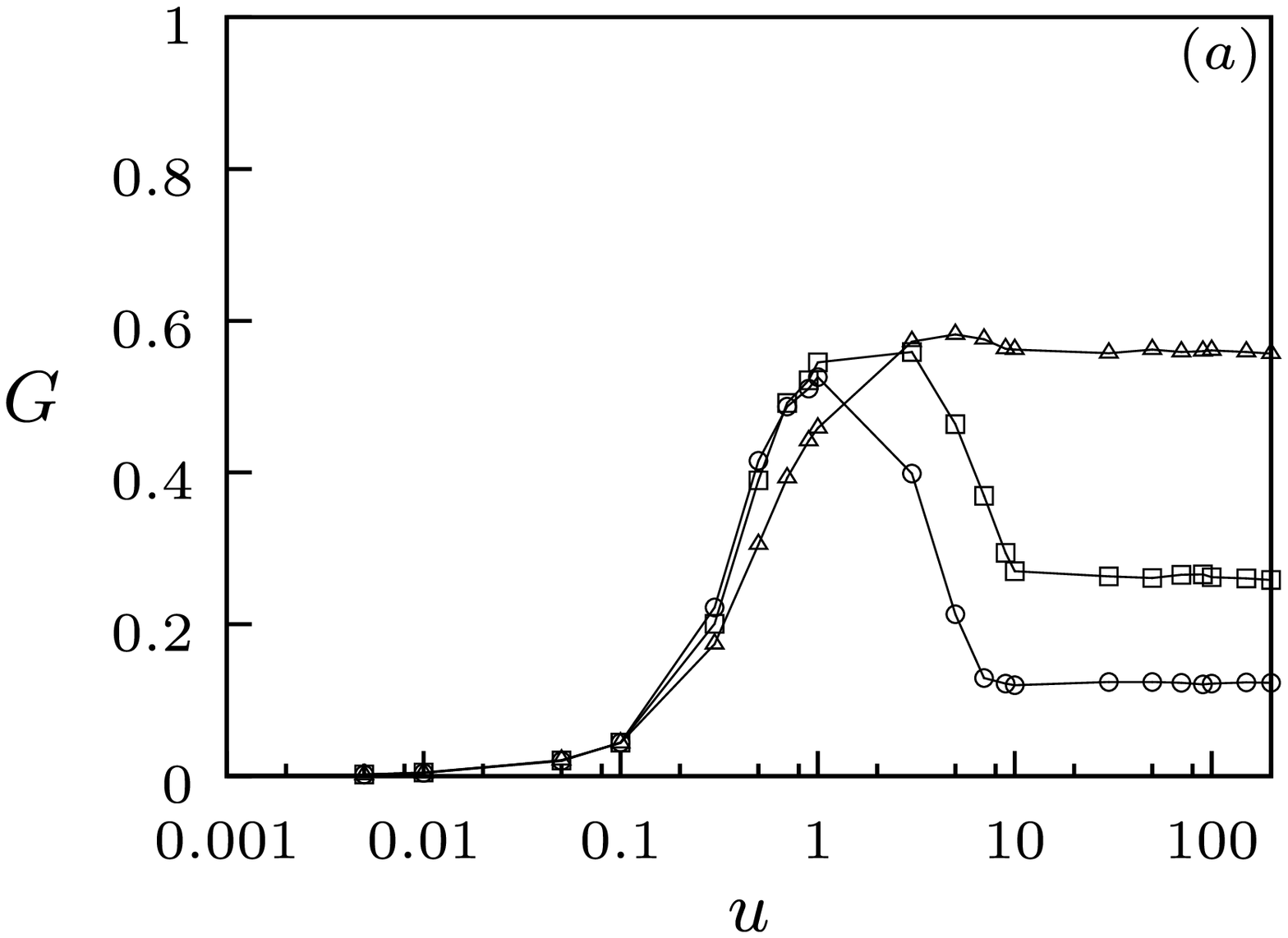}

\vspace{0.3cm}

\includegraphics[scale=0.39,angle=90]{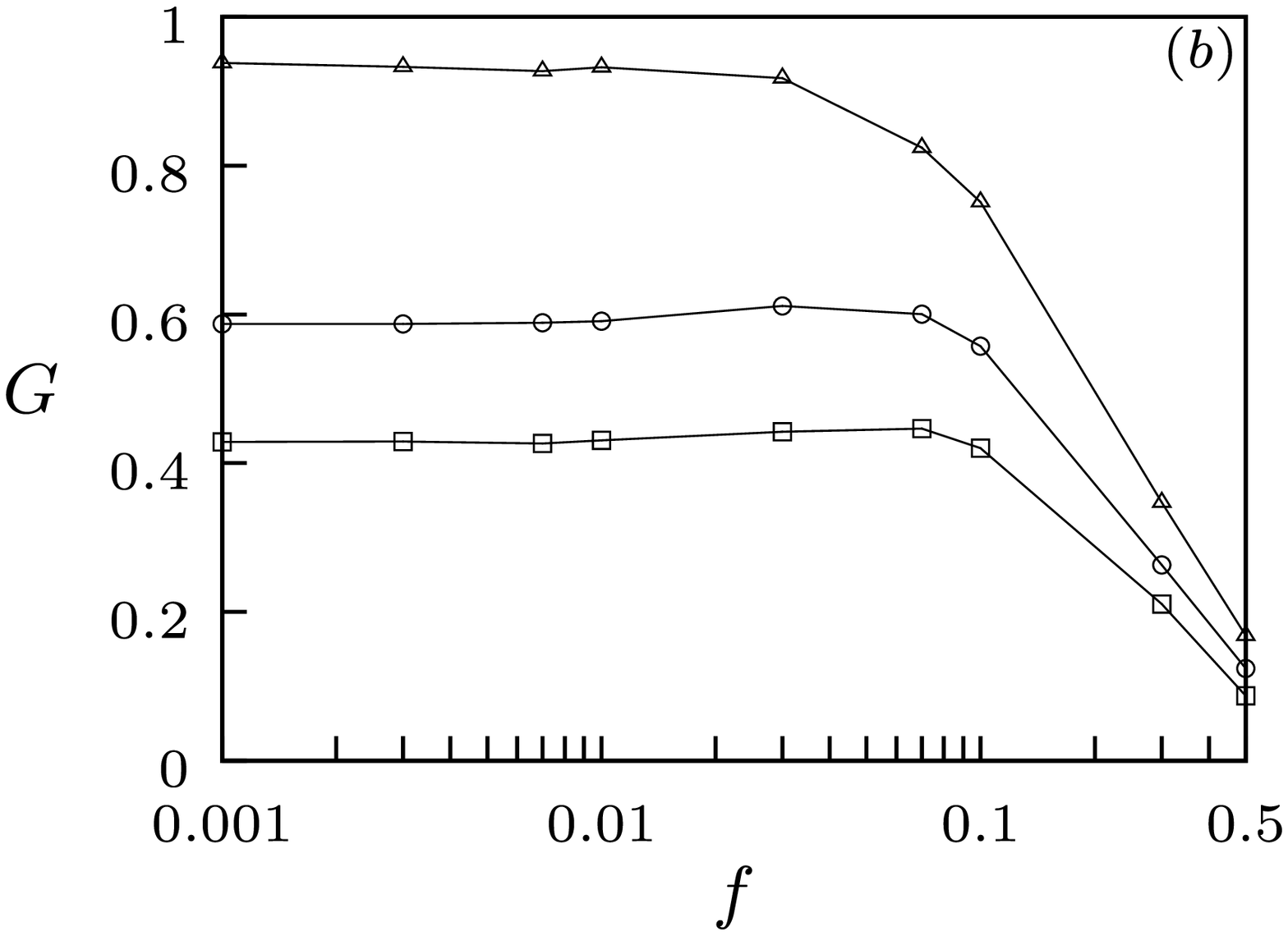}
\end{center}
\caption{Gini coefficient at $t=10^8$ as a function of $u$ with fixed $k=4$ for different $f$.
The curves correspond to $f=0.5$ (circles); $f=0.3$ (squares); and $f=0.1$ (triangles). (b) Gini coefficient at $t=10^8$ as a function of parameter $f$ for different values of $k$ and fixed $u=30$.
The curves correspond to $k=2$ (squares); $k=4$ (circles); and $k=N-1$ (triangles).}
\label{fig4} 
\end{figure}

An important variable in economic dynamics is the Gini coefficient, a statistical quantity that measures the degree of inequality in the wealth distribution in a system, defined as \cite{Gini}
\begin{equation}
G(t)=\frac{1}{2N}\frac{\sum_{i,j=1}^N |w_i(t) - w_j(t)|}{\sum_{i=1}^N w_i(t)}.
\end{equation}
A perfectly equitable  distribution of wealth at time $t$, where $w_i(t)=w_j(t), \forall i,j$, yields a value $G(t)=0$. The other extreme, where one agent has the total wealth $\sum_{i=1}^N w_i(t)$, corresponds to a value $G(t)=1$. The random, uniform distribution of wealth used as initial condition has $G(0) \approx 0$, and the average initial wealth per agent is $w_i(0)=0.5$.
Figure~\ref{fig4}(a) shows the asymptotic, statistically stationary Gini coefficient as a function of the width of the social classes $u$, for different values of the parameter $f$. 

For small values of $u$, there is a small probability of interaction between neighbors, and therefore the initial random, uniform distribution of wealth with $G\approx 0$ is maintained in the system, manifested in a low value of
 $G$. As $u$ increases, the transfer of wealth between neighbors also increases, producing a redistribution of wealth
reflected in the increase of the the Gini coefficient. A maximum of $G$ occurs around $u \approx W=1$, when each agent can initially interact with his neighbors, and therefore a greater variation with respect to the initial uniform distribution of wealth occurs in the system. For larger values of $u$, all local interactions are allowed initially. In this regime, a  redistribution of wealth should occur as the probability $f$ of favoring the poorest agents is incremented.  
This can be seen in Figure~\ref{fig4}(a) as a decrease in the values of $G$, for $u>W$, as $f$ increases.
Figure~\ref{fig4}(b) shows the Gini coefficient as a function of the probability $f$, for different sizes of 
the neighborhood $k$. The values of $G$ are almost constant for small values of $f$, but they decrease for larger values of $f$.

\begin{figure}[h]
\begin{center}
\includegraphics[scale=0.39,angle=90]{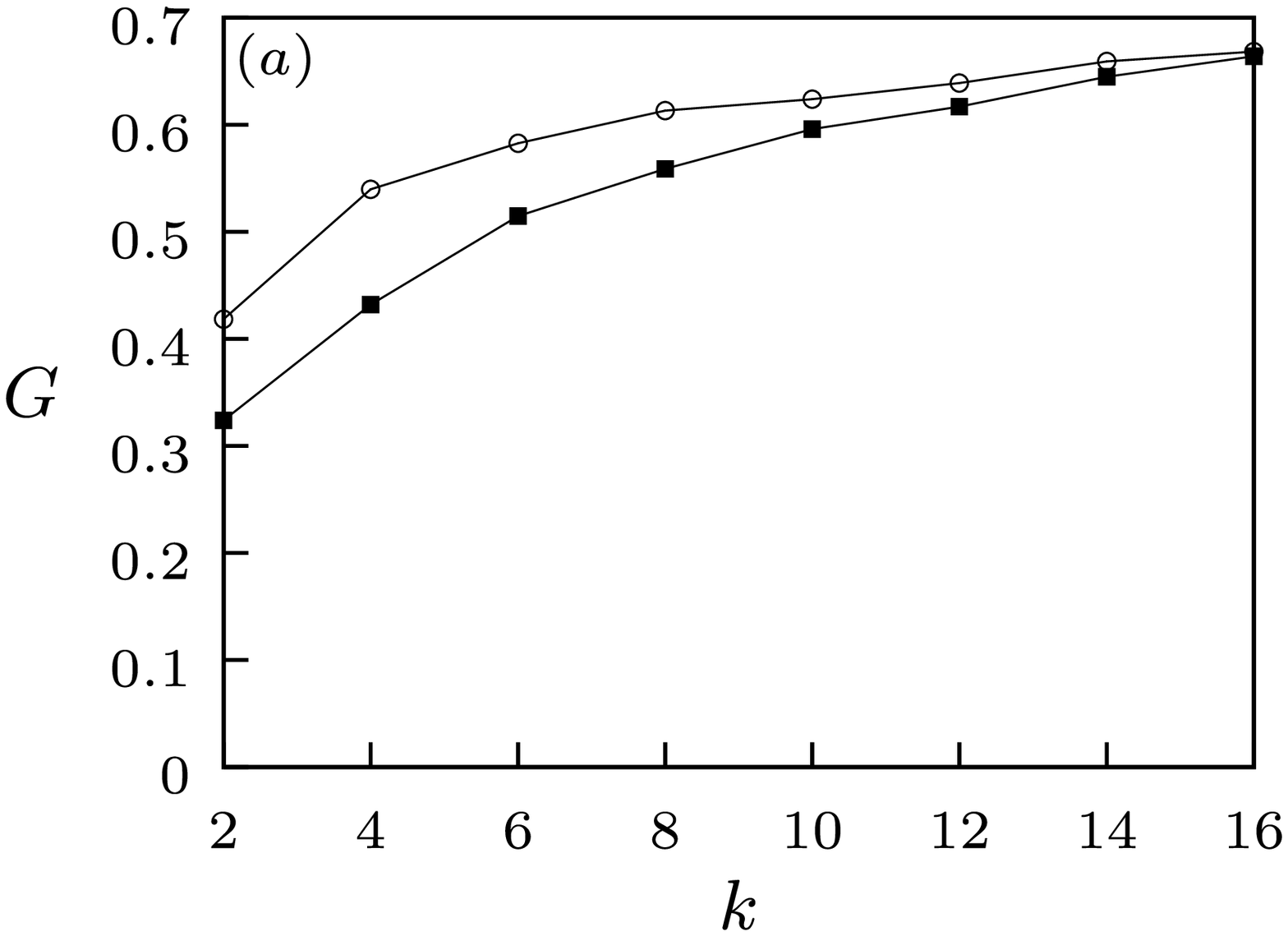}

\vspace{0.3cm}

\includegraphics[scale=0.39,angle=90]{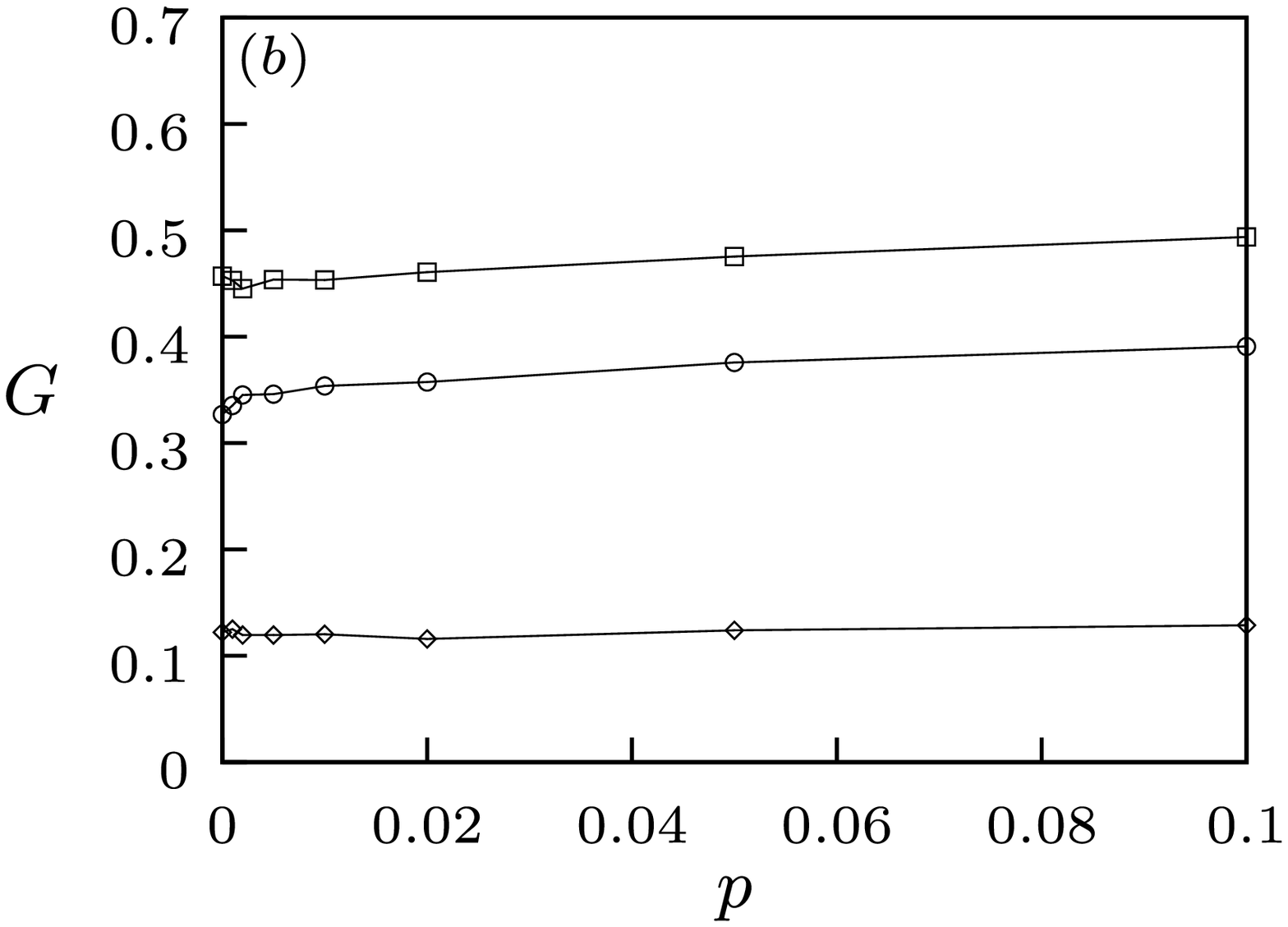}
\end{center}
\caption{(a) Gini coefficient at $t=10^8$ as a function of $k$ on a regular lattice with $p=0$, for $f=0.1$, $u=1$ (squares) and $u=10$ (circles). (b) Gini coefficient at $t=10^8$ as a function of the rewiring probability $p$ on a network with $k=4$, 
with fixed  $u=10$ and $f=0.5$ (diamonds), $f=0.3$ (circles), and $f=0.1$ (squares).}
\label{fig5} 
\end{figure}

In order to study the influence of the topology of the network on the distribution of wealth, Figure~\ref{fig5}(a) displays $G$  
as a function of $k$, on a regular lattice with $p=0$, for different values of $u$. Increasing the number of neighbors $k$ contributes to an increase in the inequality of the wealth distribution, as measured by $G$. Note that $G$ tends to an asymptotic, large value as $k \rightarrow N-1$, corresponding to a fully connected network, i. e., any agent can interact with each other in the system, losing the notion of spatial location. This corresponds to the most commonly studied situations in models of economic exchange \cite{Laguna}.

Increasing the spatial range of the interactions, represented by $k$, implies both an increment in the clustering coefficient and a decrease in the characteristic path length of the network. 
To see which of these two topological properties of the network is more relevant
for the variation of the Gini coefficient observed in Figure~\ref{fig5}(a), we plot in
Figure~\ref{fig5}(b) $G$ as a function of the rewiring probability $p$, for different values of the parameter $f$. Note that there is little change in the values of $G$ as $p$ increases, in comparison to the larger variation experienced by $G$ when $k$ is augmented in Figure~\ref{fig5}(a). The characteristic path length in the network decreases in both cases, but the clustering coefficient does not increases on the range of values of $p$ shown in Figure~\ref{fig5}(b) \cite{Watts}. Thus, the increment in the Gini coefficient observed in Figure~\ref{fig5}(a) can be mainly attributed to the increase in the clustering coefficient of the network when $k$ is varied. In other words, the size of the neighborhood is more relevant for the occurrence of an equitable distribution of wealth than the presence of long range connections in a system subject to a stratified economic exchange.

\section{Conclusions}
The inclusion of a network or a spatial location for interacting economic agents allows the use of concepts from spatiotemporal dynamical systems in economic models. We have considered a model of stratified economic exchange defined on a network and have shown that different spatiotemporal patterns can occur as the parameters of the system are varied. We have characterized these patterns as laminar, intermittent and turbulent, employing analogies from spatiotemporal dynamical systems. We have characterized the transition from a laminar state to a turbulent state through the activity of the system, that measures the average wealth exchanged in the asymptotic regime of the system. This quantity depends mainly on the dynamical parameters $u$ and $f$. Similarly, the Gini coefficient, that characterizes the inequality in the distribution of wealth, depends on the parameters $u$ and $f$. For large values of $u$, increasing $f$ increases the activity but decreases the Gini coefficient. Thus, high levels of  economic exchange activity are associated to low values of the Gini coefficient, i.e., to more equitable distributions of wealth in the system. 

The topology of the underlying network has little effect on the activity of the system $A$. In contrast, the Gini coefficient $G$ increases when the range of the interactions, represented by $k$, is increased. We have shown that the relevant topological property of the network that influences the behavior of $G$, is the clustering coefficient, instead of the characteristic path length of the network.
Figure~\ref{fig5} shows that a reduction of the Gini coefficient in a system subject to a dynamics of 
stratified economic exchange may be achieved by reducing the size of the neighborhood of the interacting agents. 

Our results add support to the view of local interactions as a relevant ingredient that can have important
consequences in the collective behavior of economic models.

\section*{Acknowledgments}
This work was supported by grant C-1692-10-05-B from Consejo de Desarrollo Cient\'ifico, Human\'istico y Tecnol\'ogico of Universidad de Los Andes,  Venezuela.
M.~G.~C. acknowledges support from  project 490440/2007-0, CNPq-PROSUL, Brazil.


\begin{thebibliography}{99}
\bibitem{Grusky} D. Grusky, Social Stratification: Class, Race, and Gender in Sociological Perspective, 3rd edition (2007) 
Westview Press.
\bibitem{Belding} T. C. Belding, arXiv:nlin/0405048v3 (2004).
\bibitem{Lehman} S. Lehmann, A. D. Jackson, B. Lautrup,  Europhys. Lett., 69 (2005)  298.
\bibitem{Vasquez} A. Vazquez, Phys. Rev. E 77, (2008) 066106.
\bibitem{Martins} T. V. Martins, T. Araujo, M. A. Santos, M. St. Aubyn, arxiv.org/abs/0809.3418 (2008).
\bibitem{Martin}  A. C. R. Martin, arxiv.org/0801.2411v2 (2008).
\bibitem{Masuda} N. Masuda,  N. Konn, J. of Theoretical Biology 243 (2006) 64.
\bibitem{Inaoka} H. Inaoka, H. Takayasu, T. Shimizu, T. Ninomiya, K. Taniguchi, Physica A 339 (2004) 621.
\bibitem{Laguna} M. F. Laguna, S. Risau Gusman, J. R. Iglesias, J. Vega, Physica A 342 (2004) 186.
\bibitem{Yakovenko} V. M. Yakovenko, in Encyclopedia of Complexity and System Science, edited by R. A. Meyers, Springer (2009). 
\bibitem{Chattarjee} A. Chatterjee, B. K. Chakrabarti, S. S. Manna, Physica A 335 (2004) 155.
\bibitem{Slanina} F. Slanina, Phys. Rev. E 69 (2004) 46102.
\bibitem{Watts} D. J. Watts, S. H. Strogatz, Nature 393 (1998) 440.
\bibitem{Barabasi} A. L. Barab\'asi, R. Albert, Science 286 (1999) 509.
\bibitem{Newman} M. E. J. Newman, A. L. Barabasi, D. J. Watts, The structure and dynamics of
networks, Princeton University Press, Princeton, N. J.,  2006.
\bibitem{Pianegonda1} J. R. Iglesias, S. Gon\c{c}alves, S. Pianegonda, J. L. Vega, G. Abramson, Physica A 327 (2003) 12.
\bibitem{Garlaschelli} D. Garlaschelli, M. I. Loffredo, J.Phys. A 41, (2008) 22.
\bibitem{Ausloos} M. Ausloos, A. Pekalski, Physica A 373 (2007) 560. 
\bibitem{Chakrabarti} A. Chakraborti, B. K. Chakrabarti, Eur. Phys. J. B 17 (2000) 167.
\bibitem{Iglesias} J. R. Iglesias, S. Gon\c{c}alves, G. Abramson, J. L. Vega, Physica A 342, (2004) 186.
\bibitem{Manneville} P. Manneville, Instabilities, Chaos and Turbulence, 2nd. edition, Imperial College Press, London, 2010.
\bibitem{Gini} M. Rodr\'iguez-Achach, R. Huerta-Quintanilla, Physica A 361 (2006) 309.
\end{thebibliography}
\end{document}